\documentclass{jpp}
\usepackage{graphicx}

\usepackage[utf8]{inputenc}
\usepackage[T1]{fontenc}
\usepackage{amsmath}
\usepackage{xcolor}

\shorttitle{Structure and dynamics of magneto-inertial rotating plasmas}
\shortauthor{Vicente Valenzuela-Villaseca et al.}

\title{Structure and Dynamics of Magneto-Inertial, Differentially Rotating Laboratory Plasmas}

\author{
    V. Valenzuela-Villaseca\aff{1}       
    \corresp{\email{v.valenzuela@princeton.edu.}}
    L. G. Suttle\aff{1},
    F. Suzuki-Vidal\aff{1}
    J. W. D. Halliday\aff{1}
    D. R. Russell\aff{1},
    S. Merlini\aff{1},
    E. R. Tubman\aff{1},
    J. D. Hare\aff{2},
    J. P. Chittenden\aff{1},
    M. E. Koepke\aff{3},
    E. G. Blackman\aff{4},
    \and S. V. Lebedev\aff{1}
}

\affiliation{\aff{1}Blackett Laboratory, Imperial College London, London SW7 2BW, UK
\aff{2}Plasma Science and Fusion Center, Massachusetts Institute of Technology, Cambridge, Massachusetts 02139, USA
\aff{3}Department of Physics, West Virginia University, Morgantown, West Virginia 26506 USA
\aff{4}Department of Physics and Astronomy, University of Rochester, Rochester, New York 14627, USA

}

\begin{document}

\maketitle

\begin{abstract}
We present a detailed characterization of the structure and evolution of differentially rotating plasmas driven on the MAGPIE pulsed-power generator (1.4 MA peak current, 240 ns rise-time). The experiments were designed to simulate physics relevant to accretion discs and jets on laboratory scales. A cylindrical aluminium wire array Z pinch enclosed by return posts with an overall azimuthal off-set angle was driven to produce ablation plasma flows that propagate inwards in a slightly off-radial trajectory, injecting mass, angular momentum, and confining ram pressure to a rotating plasma column on the axis. However, the plasma is free to expand axially, forming a collimated, differentially rotating axial jet that propagates at $\approx 100$ km/s. The density profile of the jet corresponds to a dense shell surrounding a low-density core, which is consistent with the centrifugal barrier effect being sustained along the jet's propagation. We show analytically that, as the rotating plasma accretes mass, conservation of mass and momentum implies plasma radial growth scaling as $r \propto t^{1/3}$. As the characteristic moment of inertia increases, the rotation velocity is predicted to decrease and settle on a characteristic value $\approx 20$ km/s. We find that both predictions are in agreement with Thomson scattering and optical self-emission imaging measurements.
\end{abstract}

 
\section{Introduction}
Rotating magnetized plasma flows are ubiquitous in the universe. When a plasma orbits a central object, such as a black hole or a young star, the fluid forms an accretion disc oriented perpendicular to its angular momentum vector \citep{Pringle1981}. One of the outstanding questions regarding astrophysical accretion discs is the mechanism responsible for the efficient evacuation of angular momentum needed to explain the observed luminosity of distant accreting supermassive black holes residing in Active Galactic Nuclei \citep{J.FrankA.King2002,Narayan2005}. Like-wise, protostellar and protoplanetary systems exhibit the formation of discs which feed the young star and form planets \citep{Lesur2021}. Both of these types of systems are associated with the launching of plasma jets (or winds) along the poles, which can evacuate angular momentum through magnetic torques \citep{Blandford1982,Lynden-Bell1996}. 

The typical kinematic viscosity expected on accretion discs is far too small to account for the loss of angular momentum via friction. The typical fluid Reynolds number (ratio between inertial and viscous forces) of an accretion disc is Re $\gg 1$. Under this condition, hydrodynamic turbulence can develop and potentially play a role in transporting angular momentum. \cite{Shakura1973} showed that anomalous turbulent viscosity would be enough to explain the observed accretion rate. However, a differentially rotating hydrodynamic flow with a Keplerian rotation curve is known to be robustly stable due to the Rayleigh criterion \citep{Chandrasekhar1961}, even under nonlinear perturbations \citep{Edlund2014}, which is valid when the specific angular momentum (angular momentum per unit mass) $\ell = \ell(r)$ monotonically increases with radius, such as a Keplerian system $\ell\propto r^{1/2}$.

In magnetohydrodynamics (MHD), magnetic fields can become a destabilizing agent of Rayleigh-stable flows. When the flow has an angular frequency stratification $\Omega = \Omega(r)$ that decreases with radius, the stretching of field lines between sheared layers of differential rotation drives the magnetorotational instability (MRI) \citep{Velikhov1959,Balbus1991}. The nonlinear stage of the instability triggers shear instabilities from vertically stratified `channel' flows, seeding turbulence \citep{Goodman1994,Balbus1998}. In particular, although Keplerian orbits are hydrodynamically stable since $\ell \propto r^{1/2}$, they can be MRI unstable since $\Omega \propto r^{-3/2}$. A more general flow is called quasi-Keplerian when it is Rayleigh stable but also satisfies the MRI instability condition \citep{Ji2001}. A flow with a rotation curve given by $\Omega \propto r^q$ is quasi-Keplerian for $-2<q<0$, and retains the fundamental stability properties of Keplerian orbiting disks in astrophysics, even in the absence of gravity. 

Although the MRI is commonly observed in both sheared-box and global disc numerical simulations, experimental demonstration has proven elusive. Existing laboratory experiments use the Taylor-Couette geometry to realise steady-state quasi-Keplerian, highly resistive MHD flows and study instabilities, turbulence and angular momentum transport in the nonlinear regime which develops gradually over hundreds of rotation periods. These experiments control the rotation profile from the edges of the flow, either by spinning the vessel containing a liquid, such as water or aqueous-glycerol \citep[see][]{Ji2006}, or sodium or gallium alloys under an external axial or helical magnetic field \citep[see][]{Ji2001,Goodman2002,Hollerbach2005,Stefani2006,Liu2006}; or by applying electrical currents from the edge of Hall plasmas confined by permanent magnets at the boundary \citep{Collins2012,Flanagan2020,Milhone2021}. Recently, the Princeton MRI experiment have reported the detection of the standard MRI, i.e. with a purely vertical field, using a GaInSn eustetic alloy in a Taylor-Couette apparatus \citep{Wang2022a,Wang2022b}.

High-energy-density (HED) plasmas driven at pulsed-power and laser facilities offer access to a complementary regime, different from previous experiments. These plasmas are typically hot (electron temperature $T_e > 20$ eV for pulsed-power and $T_e > 100$ eV in laser-plasma experiments, respectively) and dense, with electron number densities $n_e > 10^{18}$ cm$^{-3}$ on both cases (see reviews by, e.g. \cite{Remington2006,Lebedev2019,Takabe2021}). HED plasmas are strongly driven, exhibiting flow velocities $\mathbf{u}$ up to several hundreds of kilometers per second. As a consequence, their fluid and magnetic Reynolds numbers are much greater than unity and therefore are solidly in the ideal MHD regime \cite{Ryutov1999}. Although HED facilities were originally designed for the inertial confinement fusion program, the last decades have seen the growth of a diverse ecosystem of laboratory astrophysics experiments looking at plasma jets, supernovae explosions, collisionless shocks, magnetic reconnection, fluctuation dynamo, and others (see reviews by \cite{Remington2006,Lebedev2019,Takabe2021}). In addition, radiative cooling is often important both in this regime as well as in astrophysical accretion discs, and therefore HED laboratory experiments may open an avenue for exploring the effects of radiative cooling in a rotating plasma in conditions relevant to accretion discs which cannot be studied in liquid metal experiments.

Nevertheless, the study of rotating plasmas in the HED regime remains a largely unexplored frontier. \cite{Ryutov2011} proposed colliding multiple laser-driven plasma jets to create a rotating torus that can expel jets along the poles. This concept has inspired work on pulsed-power plasmas, including this paper. In fact, a handful of experiments investigating rotating plasmas have been fielded on pulsed-power generators (e.g. \citep{Ampleford2008,Bennett2015}) alongside with numerical simulations (e.g. \citep{Bocchi2013a,Bocchi2013b}). \cite{Cvejic2022} found self-generated rotation in an imploding gas-puff Z pinch with knife edge electrodes.  More recently, \cite{Valenzuela-Villaseca2023} reported the discovery of a quasi-Keplerian rotation curve in a dedicated laboratory astrophysics platform, which adds a crucial piece for accurate modelling of accretion disc physics. On the theory side, \cite{Beresnyak2023} found evidence of the MRI in imploding, differentially rotating plasma simulations relevant to HED experiments. Beyond physics purely of interest to accretion discs, it has been theoretically discovered that, under rotation, the compressibility and heat capacity of gases \citep{Geyko2013} and plasmas \citep{Geyko2017} can be modified.

In this article, we present a detailed experimental characterization of the plasma structure and dynamics on the Rotating Plasma Experiment (RPX) platform, extending results earlier reported by \cite{Valenzuela-Villaseca2023}. In particular, in this paper we study the wire array Z pinch ablation dynamics, plasma radial growth, structure and evolution of an axial jet launched from a rotating plasma column. Moreover, we present an in-situ determination of the flow velocity field on interferometry measurements. In addition, we present analytical models relevant to our experimental results regarding the ablating magnetic field, and evolution of the plasma radius and rotation velocity. The paper is structured as follows: Section \ref{sec:exp_setup} describes the experimental designed and diagnostics suite, together with an analytical model to calculate the ablation direction, and therefore allow controlling the total amount of initial angular momentum. Section \ref{sec:results} presents all experimental results, in particular \ref{subsec:ablation} shows that, other than the re-direction of ablation flows (which introduce angular momentum), the ablation properties are identical to standard wire array Z pinches \citep{Swadling2013,Harvey-Thompson2012a,Harvey-Thompson2012b}. In section \ref{sec:discussion} we show that the plasma exhibits a self-similar evolution, which is describable solely by conservation laws. Since our experimental results are in agreement with both our self-similar solution coupled and the calculated magnetic fields, these models can be combined to design future experiments without the need for computationally expensive numerical simulations. Finally, we use our self-similar solution to suggest future avenues for investigation.

\section{Experimental set up}\label{sec:exp_setup}

\begin{figure*}
    \centering
    \includegraphics[width=13cm]{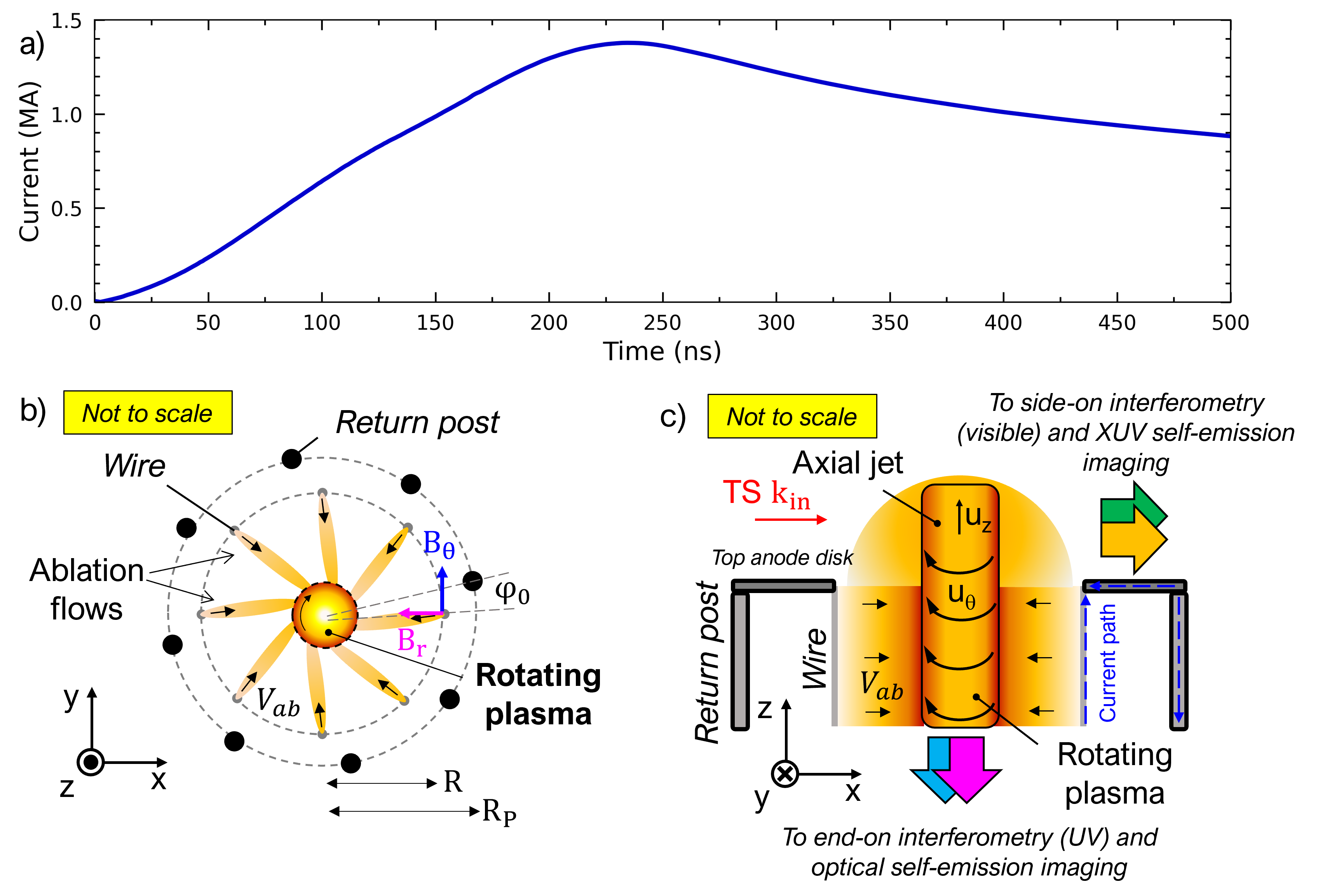}
    \caption{Schematic of experimental setup. a) MAGPIE's electrical current trace. b) End-on view of the experimental components together with the plasma dynamics in the plane of rotation. c) Side-on view. Top anode discs connects all wires and all return posts. Current path is split symmetrically through the load, but only one path through a wire and a return post is presented.}
    \label{fig:exp set up}
\end{figure*}

\subsection{The Rotating Plasma Experiment (RPX) pulsed-power platform}

The experiments were conducted on the Mega-Ampere Generator for Plasma Implosion Experiments (MAGPIE) pulsed-power generator which delivers $1.4$ MA peak electrical current with $240$ ns rise-time \citep{Mitchell1996}, as shown in Figure \ref{fig:exp set up}a. The load is a wire array Z pinch \citep{Benjamin1981,Aivazov1987,Aivazov1988,Deeney1995,Yadlowsky1996,Lebedev2000,Lebedev2001a} consisting of a $R=8$ mm radius cylindrical arrangement of eight equally spaced aluminium wires (40 $\mu$m diameter each) and $10$ mm long. The wire array is surrounded by eight 1 mm thick stainless steel return posts, offset by $\varphi_0 = 13^{\circ}$ and positioned on a $R_P=11$ mm radius cylinder, as shown in Fig. \ref{fig:exp set up}a. The return posts are used to introduce angular momentum in the experiment \citep{Bocchi2013a,Valenzuela-Villaseca2022,Valenzuela-Villaseca2023}. 

During the discharge, electrical current passes through the wires and return posts in opposite directions (Fig. \ref{fig:exp set up}b). In the first few nanoseconds, the wires heat up through Joule heating, creating a coronal plasma on their surface. The current path through the wires and return posts produces two magnetic field components $B_r$ and $B_\theta$ at each wire, which in combination with the coronal current density $\mathbf{J}$, accelerates the plasma radially inwards (by $\mathbf{J}\times B_\theta\mathbf{\hat{\theta}}$) and simultaneously introduces an azimuthal component of motion (by $\mathbf{J}\times B_r\mathbf{\hat{r}}$), producing ablation flows that follow slightly off-radial trajectories. As these ablation flows merge, they drive a rotating plasma on the axis (Fig. \ref{fig:exp set up}a).

Ablating wire array Z pinches go through distinct phases (see review by \cite{Lebedev2005}) that are relevant to the RPX. In the ablation phase, the Lorentz force transports the coronal plasma inwards and the wires remain still acting as a mass reservoir. When a significant fraction of the mass has been ablated (empirically, between $50\%$ and $80\%$), breakages along the wires appear, which quickly change the load's impedance, ending the ablation phase. During the following implosion phase, the current is suddenly redistributed and the remaining mass in the wires implodes in a shell-like implosion, snow-ploughing through the ablated trailing mass. In the stagnation phase the load's mass reaches the axis, quickly decelerating and releasing an X-ray burst (e.g., \cite{Spielman1998}). We find that RPX is able to drive rotation throughout the ablation and implosion phases. If there is a final X-ray burst from the final stagnation of the rotating plasma is currently unknown.

As shown in Fig. \ref{fig:exp set up}b, the open (vacuum) boundaries and axial confining forces allow the plasma to expand upwards producing an axial jet that rotates, transporting angular momentum.

\subsection{Diagnostics}

The plasma is probed by a suite of diagnostics which encompasses laser interferometry imaging, self-emission imaging, and Thomson Scattering that allow multiple measurements of relevant plasma parameters, such as electron density, temperature, and velocity. On each experiment, the imaging diagnostics provide two orthogonal lines of sight namely vertical (along the rotation axis, end-on, xy-plane) and lateral (perpendicular to rotation axis, side-on, xz-plane) simultaneously. Both the diagnostics probing direction and line of sight are shown schematically in Figure \ref{fig:exp set up}b and c, respectively. All diagnostics are timed relative to current start, which defines $t=0$.

An optical fast-framing camera (5 ns time resolution) images the plasma optical self-emission in the direction parallel to the rotation axis. On the same line-of-sight a two-colour Mach-Zehnder interferometry system (532 nm and 355 nm wavelengths, 0.5 ns time resolution) probes the plasma electron number density (\cite{Hutchinson2002}). The two wavelength beams are delayed by 20 ns to quantitatively probe the electron density evolution. The diagnostic provides a weighted average electron density measurement integrated along the line-of-sight $\langle n_e \rangle (x,y)Z_p = \int n_e(x,y,z) dz$  where $Z_p$ is the characteristic plasma length scale along its axis. 

A second Mach-Zehnder interferometer is aligned with the side-on field of view, perpendicular to the rotation axis. The probe beam passes above the top anode electrode to probe the axial outflows launched by the pinch. In addition, two pinhole, micro-channel plate cameras (MCP), with 5 ns time resolution, image the plasma extreme ultra-violet (XUV) self-emission. The cameras (labelled A and B) point towards the experimental axis from opposite locations. Pinhole sets are located 660 mm from the object plane and the photo-cathode front surface of the MCP at 200 mm from the pinholes. Each pinhole had a $100$ $\mu$m diameter aperture for a spatial resolution of $\sim 300$ $\mu$m. To reduce the fluence of low energy photons ($h\nu < 60$ eV), a $1$ $\mu$m Mylar filter was placed in front of the pinholes, greatly improving the image contrast. The MCP cameras provide up to 7 plasma images in each experiment which are timed to study the expansion of the plasma along the rotation axis and estimate its axial velocity $u_z$.

Quantitative measurements of electron temperature $T_e$, ion temperature $T_i$, charge state $Z$, bulk velocity in the plane of rotation $\mathbf{u} = (u_x, u_y)$ are obtained using an optical Thomson Scattering (TS) diagnostic (\cite{Swadling2014,Suttle2021}). A laser beam (532 nm wavelength, 3 J, 7 ns) is focused across the experimental axis with the incident wavevector $\mathbf{k_{in}}$ perpendicular to it. The beam passes in the horizontal direction approximately $5.5$ mm above the upper surface of the top anode electrode, and therefore probes the axial outflows. The scattered spectra is collected by two linear arrays of 14 optic fibres each were located in the horizontal plane (perpendicular to the rotation axis) at $\pm 90^{\circ}$ relative to $\mathbf{k_{in}}$, labelled bundles A and B. The fibre bundles are coupled to an imaging spectrometer to measure the TS spectra from the collection volumes. Thus, the combined TS configuration is sensitive to the orthogonal scattering vectors $\mathbf{k_{A(B)}} = \mathbf{k_{out, A(B)}} - \mathbf{k_{in}}$, where $\mathbf{k_{out, A(B)}}$ are the directions of collection for bundle A (and B, respectively).

The TS diagnostic probes local plasma conditions. A 532 nm wavelength, 7 ns, 3 J laser beam is focused through the plasma. The $200$ $\mu$m beam waist defines the spatial resolution along the collection direction. The spatial resolution along $\mathbf{k_{in}}$ is given by the product of the magnification $\sim 0.54$ and the diameter of each optic fibre ($100$ $\mu$m), yielding an imaged collection volume of diameter $\sim 180$ $\mu$m. Therefore, the diagnostic samples plasma conditions averaged over a collection region with volume $\sim 200\times \pi \times180^2/4$ $\mu$m$^3$ region. The characteristic collection diameter is much smaller than the characteristic plasma-gradient length scales. The spectral features can be fitted using the dynamical form factor $S(k,\omega)$ for ion-acoustic wave TS (\cite{Froula2011,Hare2017Thesis}). The fitting process was constrained by using the electron density measured by the side-on interferometry diagnostic to reduce the number of fitting variables. 

\begin{figure*}
    \centering
    \includegraphics[width=11cm]{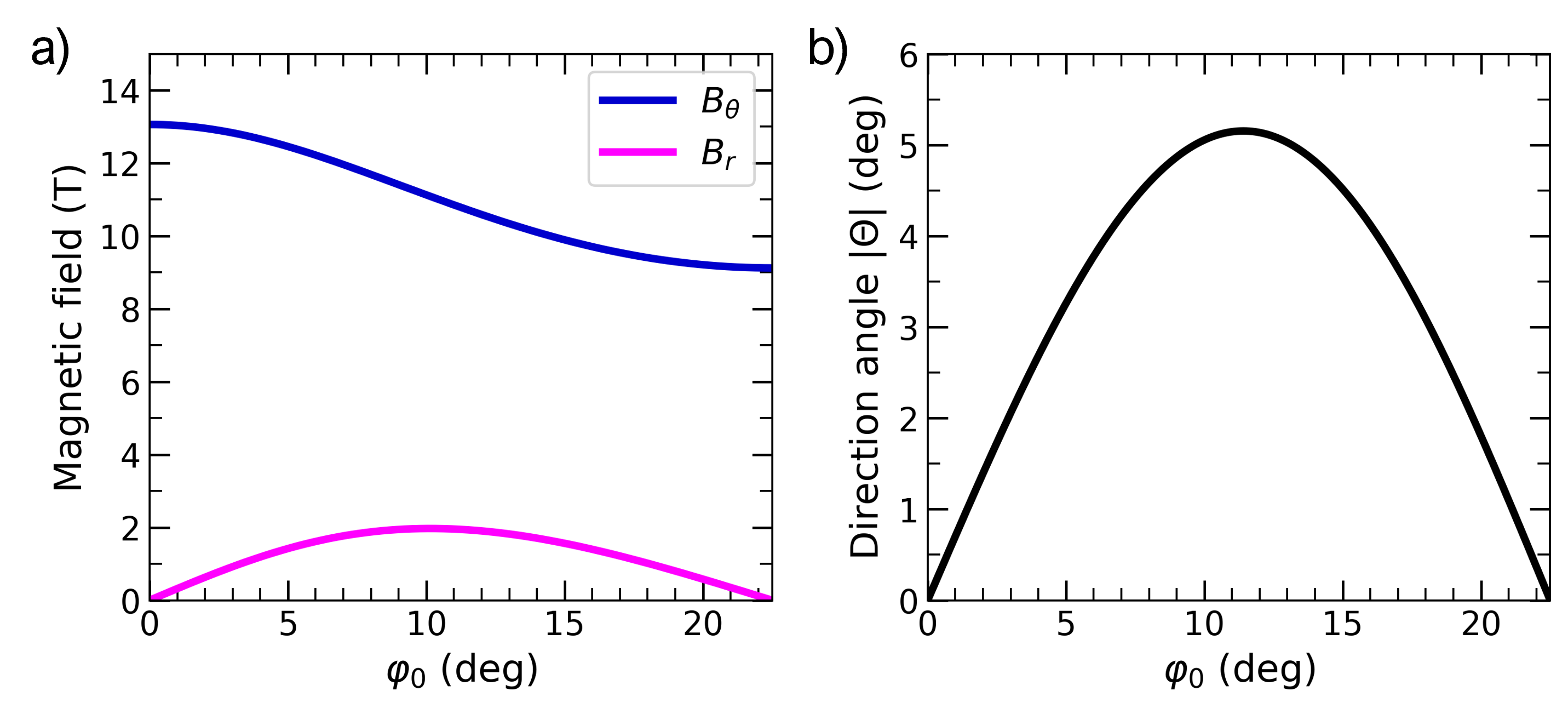}
    \caption{Magneto-static model calculated for relevant configuration: $R=8$ mm, $R_P=11$ mm, and $I=1.4$ MA. a) Magnetic field components at wire position. b) Direction angle with respect to radial vector.}
    \label{fig:vacuum fields}
\end{figure*}

\subsection{Magnetic field configuration and direction of ablation flows}\label{subsubsec:vacuum eqs}

The ablation direction of the streams can be calculated analytically from a magneto-static model.  In cylindrical coordinates and without loss of generality, the magnetic field components evaluated in the wire located at $r=R, \theta = 0$ are given by Biot-Savart's law (\cite{Felber1981})

\begin{align}
B_r(R) &= -\frac{\mu_0 I}{2\pi N} \sum_{j=0}^{N-1}\frac{R\sin(\varphi_0 + 2\pi j/N)}{R^2+R_P^2-2RR_P\cos(\varphi_0 + 2\pi j/N)},  \label{eq:radial}\\
B_\theta(R) &= \frac{\mu_0 I}{2\pi N} \left[ \frac{N-1}{2R} - \sum_{j=0}^{N-1}\frac{R-R_P\cos(\varphi_0 + 2\pi j/N)}{R^2+R_P^2-2RR_P\cos(\varphi_0 + 2\pi j/N)} \right], \label{eq:azimuthal}
\end{align}

where $R$ is the wire array radius, $R_P$ is the return post array radius, $\varphi_0$ is the wire-post off-set angle, $N$ is the number of wires (and return posts), $I$ is the electrical current. The equations above describe the magnetic field components generated by all other wires together with all return posts.

For the cases $\varphi_0 = m\pi/8$ ($m$ an integer), equation (\ref{eq:radial}) becomes zero, and for $\varphi_0 = m\pi/4$ equation (\ref{eq:azimuthal}) is reduced to the standard wire array Z pinch vacuum field. Hence, for any other value of $\varphi_0$, angular momentum will be introduced because the resulting $\mathbf{J}\times\mathbf{B}$ force has a radially inwards and an azimuthal component. The ablation flow is ejected along an off-radial trajectory parameterized by the direction angle $\theta \equiv \tan^{-1}(B_r/2B_\theta)$. For the wire and return post array dimensions used in the experiment, Fig. \ref{fig:vacuum fields}a shows the vacuum fields calculated for parameters relevant to these experiments. They show that the radial magnetic field component is approximately $10\%$ of the azimuthal component. The result also shows that the azimuthal component changes slightly due to the return posts.

Fig. \ref{fig:vacuum fields}b shows the off-radial direction angle $\theta$ as a function of $\varphi_0$. They show that the maximum direction angle is obtained for $\varphi_0 = 11.25^{\circ} = 119.35$ mrad, as expected from geometric arguments presented above. The expected deflection in the experiment is in the range $\theta \sim 4^{\circ} - 5^{\circ}$ for the array parameters described above.


\section{Experimental results}\label{sec:results}

\subsection{Ablation dynamics and formation of rotating plasmas}\label{subsec:ablation}

\begin{figure*}
    \centering
    \includegraphics[width=13cm]{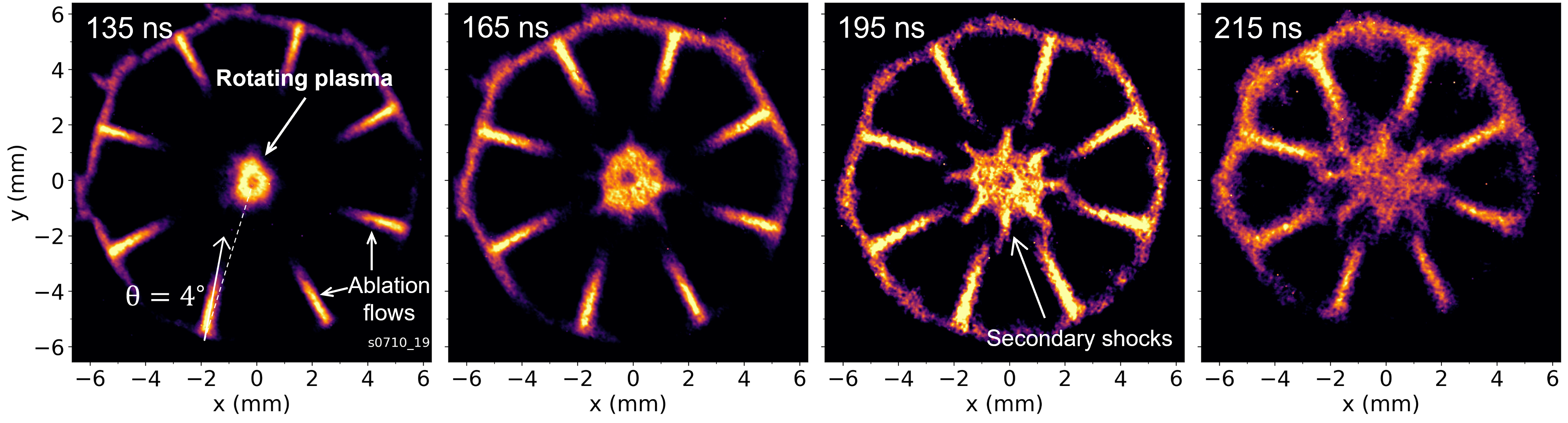}
    \caption{Sequence of end-on optical self-emission images obtained in a single experiment. Emission from the outer edges are likely produced by some residual ablation of the experimental hardware. The self-emission intensity along the ablation flows decreases due to the combination of radiative cooling effects and a decreasing density towards the axis \cite{Valenzuela-Villaseca2023}.}
    \label{fig:optical self emission}
\end{figure*}

Fig. \ref{fig:optical self emission} shows a sequence of optical self-emission images of the formation and evolution of the ablation flows and rotating plasma. Since the plasma contains aluminium ions which are not fully stripped (charge state $Z\sim 7$) and emit copious optical light through atomic transitions, the intensity of the images is a complicated function of the density and temperature of the plasma, but can be used for qualitative evaluation of the plasma structure. The images show elongated features oriented inwards. These are the ablation flows ejected from each of the aluminium wires. The analysis shows that their direction is not along a radial direction, but rather tilted azimuthally by $\theta = 4^{\circ}\pm 1^{\circ}$ which is consistent with the theoretical prediction of the vacuum magneto-static model presented in Section \ref{sec:exp_setup}. The self-emission of the ablation flows diminishes as they propagate inwards due to radiative cooling effects.

On the axis, a circular bright ring is formed at around $130$ ns, corresponding to the rotating plasma. This formation time can be used to estimate the ablation velocity $V_{ab}\sim 70$ km/s, consistent with direct Thomson scattering measurements by \cite{Harvey-Thompson2012a} for standard wire array Z pinches. The self-emission structure exhibits reduced intensity on-axis. This is consistent with a dense plasma shell surrounding a low-density core. This has been observed in previous experiments by \cite{Bennett2015} also and is produced by the centrifugal barrier associated with dynamically significant angular momentum \citep{Valenzuela-Villaseca2023b}. Unfortunately, this sequence of images does not allow to estimate the rotation velocity, since the smooth ring characterizing the plasma has no discernible features that co-rotate with the plasma which can be tracked with time. At $\sim 170$ ns, radially-oriented features emerge from the ring. These are secondary shocks produced by the oblique interaction of the ablation flows undergoing thermal expansion \citep{Swadling2013}.

\begin{figure*}
    \centering
    \includegraphics[width=13cm]{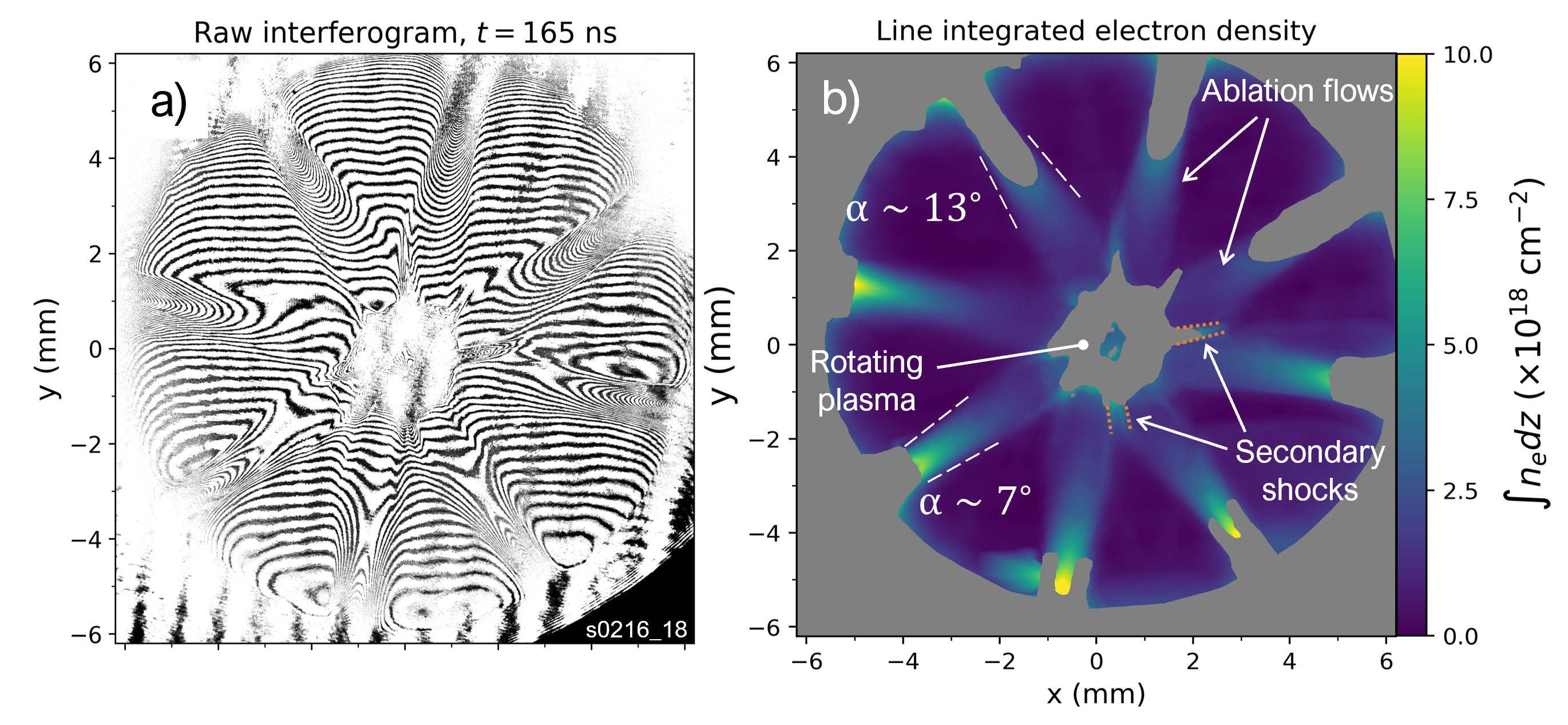}
    \caption{End-on interferometry results obtained from the 355 nm wavelength probe. a) Raw interferogram. b) Processed line-integrated electron density map. The contour of two secondary shocks have been highlighted with orange dotted lines.}
    \label{fig:end on interferometry}
\end{figure*}

 Quantitative measurements of the density structure of the plasma at a given time can be obtained through laser interferometry. Figure 4a shows a laser interferogram obtained after the formation of the rotating plasma. The bright and dark fringes correspond to regions of constructive and destructive interference, respectively, between the probe beam (passing through the plasma) and a reference beam (passing around the plasma). The interferometer is aligned to have the reference interferogram (i.e. in the absence of plasma) with horizontal fringes. Thus, any bending or deformation of the interferogram is due to the added refractive index, which is proportional to the plasma electron density. Figure 4b shows the inferred line-integrated electron density map calculated using the MAGIC2 code \citep{Swadling2013,Hare2019}, which constructs a phase shift map by interpolating between the fringes and then comparing the reference and shot interferograms. The phase shift map is proportional to $\int n_e dz$, hence calculating a map of the electron density map integrated along the line-of-sight. Since, in the image there is no region with zero fringe shift, the diagnostic can only relative changes in electron number density. For clarity, we have defined the region with minimum phase shift as having zero density (which coincides with the regions between the ablation flows). To unfold the (volumetric) electron density we must assume some symmetry. For the end-on view, the plasma can be assumed to be homogeneous along its length $L$ if we neglect contributions of the axial jet ejected from this region, which is significantly lower density than the plasma inside the wire array \citep{Valenzuela-Villaseca2023}. The plasma density is dominated by the region between the two electrodes (the axial jets are one order of magnitude less dense, which will be shown in the next section). Therefore, L = 10 mm corresponding to the array’s length.

The data show the interaction between the inwards propagating ablation flows as they reach the rotating plasma column. The electron density decreases radially inwards by a factor of $\sim 10$, which may help explain features in the optical self-emission images. The ablation flows remain extremely collimated during the experiment, with divergence angles in the range $\alpha = 10^{\circ} \pm 3^{\circ}$. This can be used to calculate their sonic Mach number
\begin{equation}
    M_s = \frac{u}{c_s} = \frac{1}{\sin{\alpha/2}} = 10\pm 3,
\end{equation}
where $u$ is the flow velocity ($=V_{ab}$) and $c_s$ is the plasma sound speed. A characteristic plasma temperature $T=10$ eV can be inferred by considering the ion sound speed $c_s \sim \sqrt{k_BT/m_i}$, where $m_i$ is the ion mass. Our results are consistent with previous reports by \citep{Harvey-Thompson2012a,Harvey-Thompson2012b, Swadling2013}. We note that Thomson scattering measurements revealed $T_i$ and $T_e$ to be $>10$ eV, which indicates the the ablation flows are magnetically collimated by diverting part of the driver's current through the streams \citep{Harvey-Thompson2012a,Swadling2013}. Since our results exhibit similar structure, we conclude that the effect of the radial magnetic field applied mainly re-directs the plasma streams, rather than changing the overall ablation properties of the wire array Z pinch. Experimentally, we find that the ablation flows can sometimes break from axisymmetry, likely due to differences in the ablation azimuthal divergence can be ascribed to discrepancies in the wire ablation rate due to uneven current distribution during the electrical discharge.

The interferogram also helps visualizing the origin of the secondary shocks. As the ablation flows thermally expand in the azimuthal direction, they can interact with their immediate neighbours before merging with the rotating plasma on the axis. This produces oblique, standing shocks which remain stationary due to the steady state ablation dynamics of the Z pinch. 

The fringes can be seen in the inflows but are lost in the outer region of the rotating plasma. This is due to density gradients perpendicular to the line of sight deflecting the probe beam out of the optical system’s acceptance angle, as pulsed-power plasmas have typical densities below critical density for 355 nm wavelength (Nd:YAG 3rd harmonic) and the effect of absorption in negligible. As a consequence, the electron density of the rotating plasma cannot be estimated from end-on probing. However, there are visible interferogram fringes in the region on axis. This is consistent with either a density distribution with an on-axis flat profile (i.e. no density gradients) or a hollow density structure wide enough to allow the probe to cross the plasma.

\subsection{Propagation and density structure of axial jets}\label{subsec:axial_jets}

\begin{figure*}
    \centering
    \includegraphics[width=13cm]{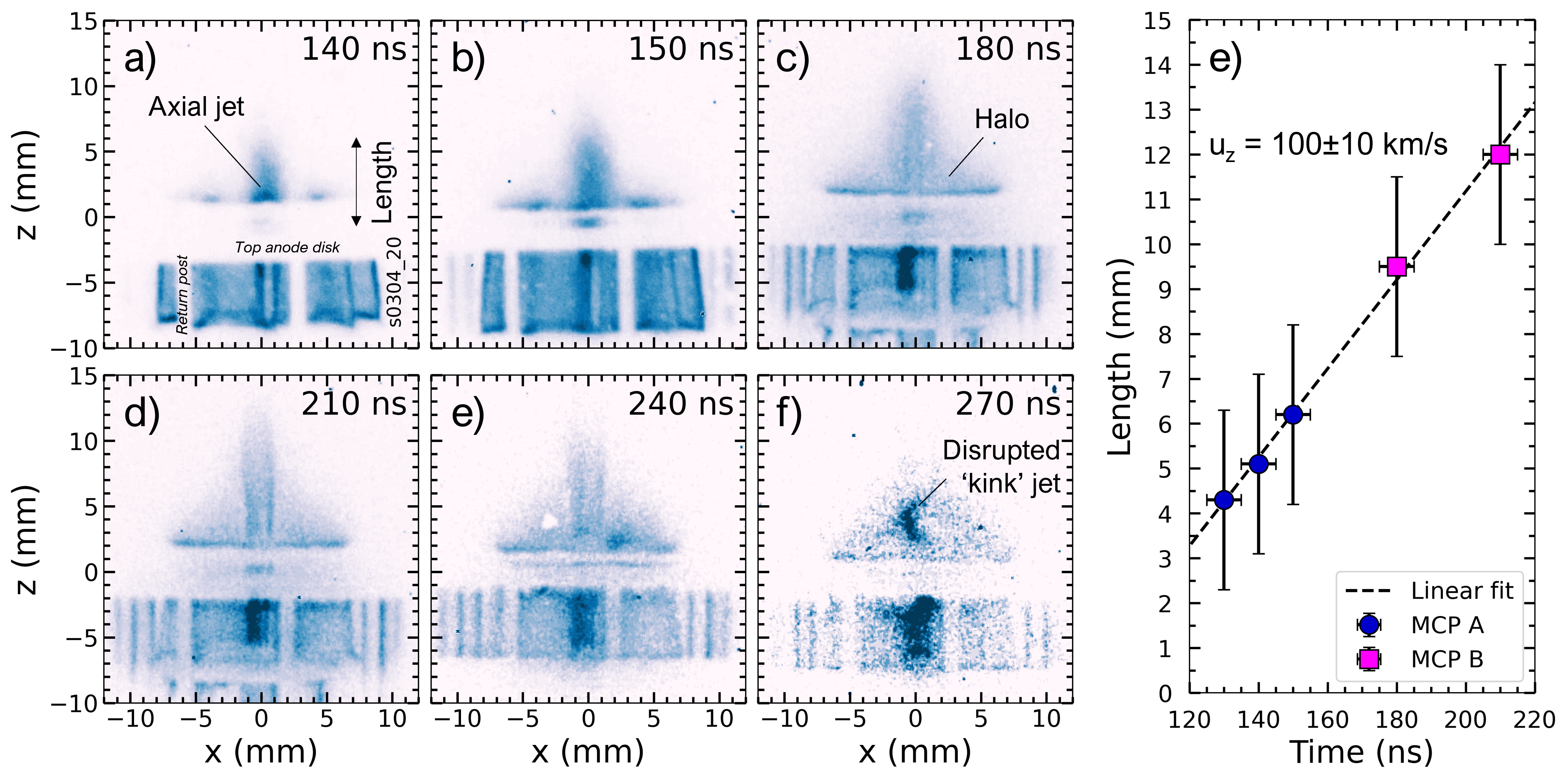}
    \caption{Side-on XUV plasma self-emission data. a$-$f) Sequence of pinhole images obtained in the same experiment. Features corresponding to hardware components are annotated in italics. Panel (d) corresponds to the image presented in (\cite{Valenzuela-Villaseca2023}), reprinted with the author's permission. e) Length of axial jet, measured using the full height at half maximum intensity, as a function of time obtained as measured from the MCPs (labelled A and B) located on two opposing lines of sight. Image at $t=130$ ns is not presented and length corresponding to $t=240$ ns was not used in linear fit.}
    \label{fig:XUV_sequence}
\end{figure*}

Panels a--f in Fig. \ref{fig:XUV_sequence} shows a sequence of extreme ultra-violet (XUV) images of the launching and propagation of the axial jet. A $1$ $\mu$m-thick Mylar filter was used to remove soft XUV emission and improve the contrast. In these images, dark regions correspond to high plasma XUV self-emission intensity, whereas the light-coloured regions correspond to non-emitting metallic hardware and vacuum regions. The z-axis is oriented vertically with the position z = 0 defined as the approximate location of the top disc’s upper surface. Therefore, the wire array plasma is restricted to the region z < 0. Vertical light-coloured strips which correspond to the shadows of the return posts. The results show that the posts remain static throughout the experiment, with very little visible ablation and no overall deformations. This robustness is key to sustain a steady state off-axis ablation.

The sequence reveals the presence of plasma on the region z > 0. Since the initial experimental set up does not accelerate plasma axially (as opposed to, e.g. \citep{Ampleford2008}), any outflows reaching this region must have been launched upwards by the wire array plasma through gradients of thermal and/or magnetic pressure. The outflows can be described as consisting of a well collimated axial jet on the axis, and an unstructured plasma halo which surrounds the axial jet. The images show that the plasma  remains in a quiescent state, with no large perturbations or significant changes in structure morphology up until $t = 270$ ns, both in the wire array and outflow regions. This sets a $\sim 140$ ns time-frame over which rotation can be reliably sustained, as the plasma is first formed at $\sim 130$ ns after current start. This shows that rotation can be maintained through the ablation and implosion phases of the Z pinch.

The axial velocity of the flow can be estimated by following the upper end of the jet, which is representative of its length, through the sequence of images for the two XUV cameras fielded in the experiment. The result is presented in Fig. 5b, with each XUV camera presented in different colours. The horizontal uncertainty is given by the 5 ns of time integration, whereas the vertical error bars estimate the uncertainty in the position of the tip of the jet due to the smooth extinction of the plasma self-emission. The results show the linear trend of the position of the jet with time, from which a vertical velocity $u_z  = 100 \pm 10$ km/s can be estimated. 

\begin{figure*}
    \centering
    \includegraphics[width=13cm]{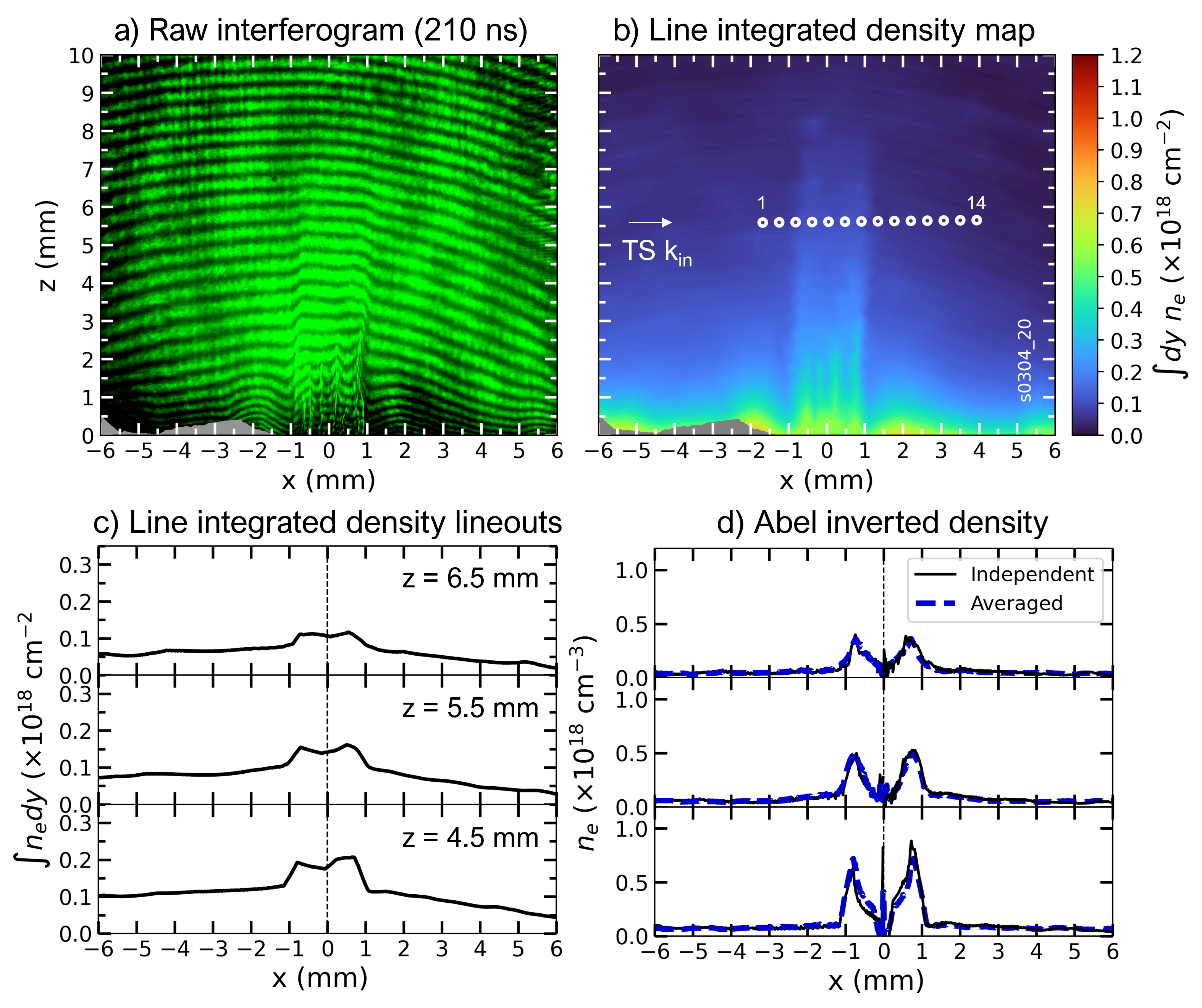}
    \caption{Side-on laser interferometry results (532 nm, 0.5 ns time-resolution) at 210 ns after current start. a) Raw interferogram. b) Line-integrated electron density map. TS probe incident wave-vector points from left to right as indicated by the white arrow and collection volumes are shown in white circles. c) Density lineouts at three heights. Middle panel corresponds to a lineout at the height of TS probe. d) Abel inverted density profiles unfolded from lineouts on panel c. Inversion algorithm assumes axisymmetry independently on both sides and the result is presented in solid black lines. An average between left- and right-hand-side inversion is shown in a blue dashed arrow.}
    \label{fig:side on interferometry}
\end{figure*}

The density structure of the axial jet was studied using laser interferometry measurements. Fig. \ref{fig:side on interferometry}a shows a typical side-on raw interferogram. The coordinate system is consistent with FIG 5, i.e. the line z = 0 corresponds to the upper surface of the top electrode. Fig. \ref{fig:side on interferometry}b shows the line-integrated electron density map inferred from panel (a). The central elongated structure corresponds to the axial jet. This result confirms the high-aspect-ratio ($H/R \sim 10$, where $H$ is the jet length and $R$ its radius) and collimation degree of the jet ($\sim 3^{\circ}$ divergence angle). The jet’s base exhibits the formation of vertical filaments. These are consistent with the secondary shocks formed by the oblique collision of adjacent ablation flows, which can expand axially upwards and become present in the outflow region. 

Lineouts of Abel-inverted (volumetric) electron density along at constant height were calculated using a peeling-onion method \citep{Dasch1992,Valenzuela-Villaseca2022}. This technique requires assuming cylindrical symmetry of the plasma, and therefore it is only applicable to the regions $z > 3$ mm, above the filamentary secondary shock structures that break the jet’s symmetry.

Figs. \ref{fig:side on interferometry}c and d summarize the results obtained at three different heights. The line integrated density lineouts (panel c) show that the jet is not exactly axisymmetric. To estimate the effect of the asymmetry in the result, we calculated the Abel inversion separately on the left- and the right-hand side about the axis. We show the results on black lines on (panel d). The two inversions are averaged and shown in blue. Hence the difference between the one-sided inversion and the averaged inversion can be used to estimate the uncertainty introduced by the assumed axisymmetry. The results show that the effect is $< 10\%$.

The measurement reveals the hollow density structure of the plasma, as it exhibits a steep density decrease towards the axis. This is consistent with the observed optical self-emission profile discussed above, i.e. a dense shell surrounding a lower density core. This density profile is present at all values of z, and therefore we conclude that it is retained as the axial jet propagates upwards, although we observe a slight decrease in the average density gradient from $<dn_e/dr>=0.5\times 10^{18}$ cm$^{-3}$/mm at z = 5.5 mm to $<dn_e/dr>=0.3 \times 10^{18}$ cm$^{-3}$/mm at z = 6.5 mm. The maximal electron density of the axial jet gradually decreases with height at an average gradient $<dn_e/dz>=0.1\times 10^{18}$ cm$^{-3}$/mm, likely due to the ablation time history imprinted by the driver’s discharge. The radial location of maximal electron density does not change on the lineouts, which further demonstrates the collimation of the jet, with negligible mass diffusion down the density gradients. 

\subsection{Thomson Scattering measurements of plasma flow velocity and temperature profile}\label{subsec:velocity and temperature}

The absence of inwards mass diffusion is consistent with the axial jet’s density profile being supported by a centrifugal barrier, which implies that it is spinning \citep{Valenzuela-Villaseca2023b}. In the experiments, direct measurements of the rotation and radial flow velocity components were obtained using an optical Thomson Scattering (TS) diagnostic. The TS probe beam passes horizontally through the rotating jet at $z = 5.5$ mm above the top disc’s upper surface and passing approximately through the plasma rotation axis. We extend the analysis reported by \cite{Valenzuela-Villaseca2023}. In particular, we measure from the orthogonal interferometry line of sight the impact parameter which is important to interpret the flow velocity components as azimuthal and radial.

\begin{figure*}
    \centering
    \includegraphics[width=13.5cm]{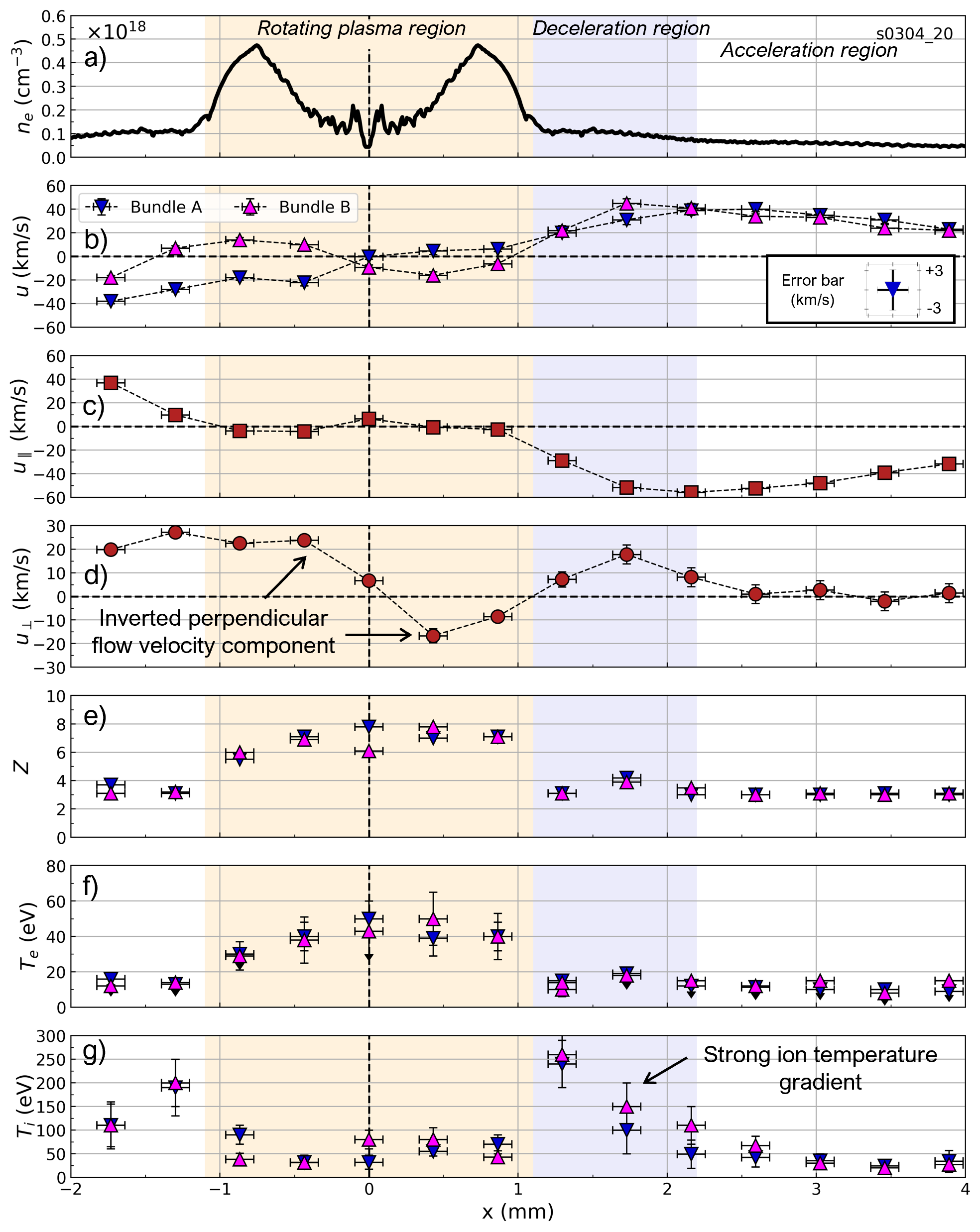}
    \caption{Inferred plasma parameters obtained for plasma outflows (axial jet and halo). Regions of interest are highlighted in background colour: rotating plasma region in orange, and deceleration region in blue. a) electron density obtained from interferometry at the height of interest. b) Horizontal flow velocity components along scattering vectors. Inlet: characteristic Doppler shift error bar. Horizontal error bars given by TS collection volume diameter (200 $\mu$m). c) Plasma velocity component along laser wave-vector $\mathbf{k_{in}}$ (flow radial velocity).  d) Plasma velocity component perpendicular to laser wave-vector (flow rotation velocity). e) Average charge state $Z$. f) Electron temperature. Downwards-pointing arrow indicates value is an upper constraint (single-peaked IAW spectra). g) Ion temperature.}
    \label{fig:Thomson}
\end{figure*}

Fig. \ref{fig:Thomson} summarizes the plasma parameters measured from TS, together with the electron density lineout inferred from laser interferometry measurements (FIG. 8a) at the height of interest. The position of the plasma axis projected along the line-of-sight is shown on all panels. The plasma is inhomogeneous and can be separated in 3 distinct regions: the rotating plasma region (- 1.2 mm < x < 1.2 mm), a deceleration region (1.2 mm < x < 2 mm), and an acceleration region (x > 2 mm). The latter two correspond to sub-regions in the plasma halo indicated in Fig. \ref{fig:XUV_sequence}.

The TS spectrum is sensitive to the flow velocity components projected on the scattering vectors $(\mathbf{k_A}, \mathbf{k_B})$ defined in Section \ref{sec:exp_setup}. The Doppler shifted velocity components obtained from the fit on both fibre bundles are shown in Fig. \ref{fig:Thomson}b. To interpret the measurement, we reconstruct the radial and azimuthal flow velocity components using the Cartesian $\mathbf{k_A}$ and $\mathbf{k_B}$ vectors. . Since the observation points lie on the horizontal plane (perpendicular to the rotation axis), then the velocity components parallel and perpendicular to $\mathbf{k_{in}}$ are
\begin{align}
    u_{\parallel} = -\frac{1}{\sqrt{2}}\left(u_A + u_B \right)&, \quad
    u_{\perp} = \frac{1}{\sqrt{2}}\left( u_A - u_B \right),    \label{eq:velocity components}
\end{align}
where $(u_{A},u_{B})$ are the flow velocity components projected on the scattering vector basis. When the probe beam passes through the rotation axis, the vector basis $(u_{\parallel},u_{\perp})$ correspond to the radial and azimuthal flow velocity components\footnote{Up to a $\pm$ sign determined by the transformation from the fixed vector basis $(\mathbf{k_A},\mathbf{k_B})$ to co-rotating of polar coordinates.}. For clarity, the following description we assumes this is the case, but below we will apply a correction to a small offset between the probe beam and the rotation axis (Section \ref{subsec:impact parameter}). 

Fig. \ref{fig:Thomson}c shows the flow velocity component parallel to $\mathbf{k_{in}}$ (i.e. radial velocity). The data show that the halo flows inwards and merges into the axial jet. In the outermost (acceleration) region, the plasma accelerates from $40 \pm 3$ km/s to $60 \pm 3$ km/s. However, approximately 1 mm before reaching the axial jet (i.e. before the density jump associated with the rotating plasma shell), the inflow decelerates to $25 \pm 4$ km/s. This deceleration can also be seen on the left-hand-side of the lineout ($x < -1.2$ mm). Inside the rotating plasma region, the radial velocity is approximately zero, with a maximum value $5 \pm 3$ km/s. This indicates that the plasma is in an approximate equilibrium state, with negligible mass radial transport.

The rotation of the axial jet is manifested in the consistent change of sign of $u_\perp$ about the axis, shown in Fig. \ref{fig:Thomson}d. In the rotating plasma region, the data points on the left of the axis have a consistent positive sign, whereas those on the right have a negative one. This is a direct measurement of rotation. The point on the axis has non-zero $u_\perp$ due to the slight misalignment of the probe beam with the rotation axis (see Section \ref{subsec:impact parameter}). The maximum rotation velocity $23 \pm 3$ km/s is observed at $x=-0.4$ mm. In the deceleration region, $u_\perp$ does not exhibit a monotonic behaviour. This indicates that this velocity component is associated with the azimuthal expansion of the inflows (shown in Fig. \ref{fig:end on interferometry}b), rather than rotation. The acceleration region shows negligible flow perpendicular to the laser beam.

In the halo, the IAW spectral feature is single peaked, indicating that $3T_i > ZT_e$ \citep{Froula2011}, which means that the fitting only provides an upper constraint on the electron temperature. The average charge state and electron temperature remain approximately constant in the halo (Fig. \ref{fig:Thomson}e and f), with $Z \sim 3$ and $T_e \leq 15$ eV. Inside the rotating plasma, the electron temperature increases towards the axis, from $T_e = 30 \pm 12$ eV to $45 \pm 15$ eV. The average charge state is calculated using a look-up table generated from the atomic kinetic code SpK \citep{Hare2017Thesis,Crilly2023} which calculates ionization states in the relevant conditions to decouple the product $ZT_e$. The result is a relatively constant charge state $Z \sim 7$ across the jet.

\begin{figure*}
    \centering
    \includegraphics[width=13.5cm]{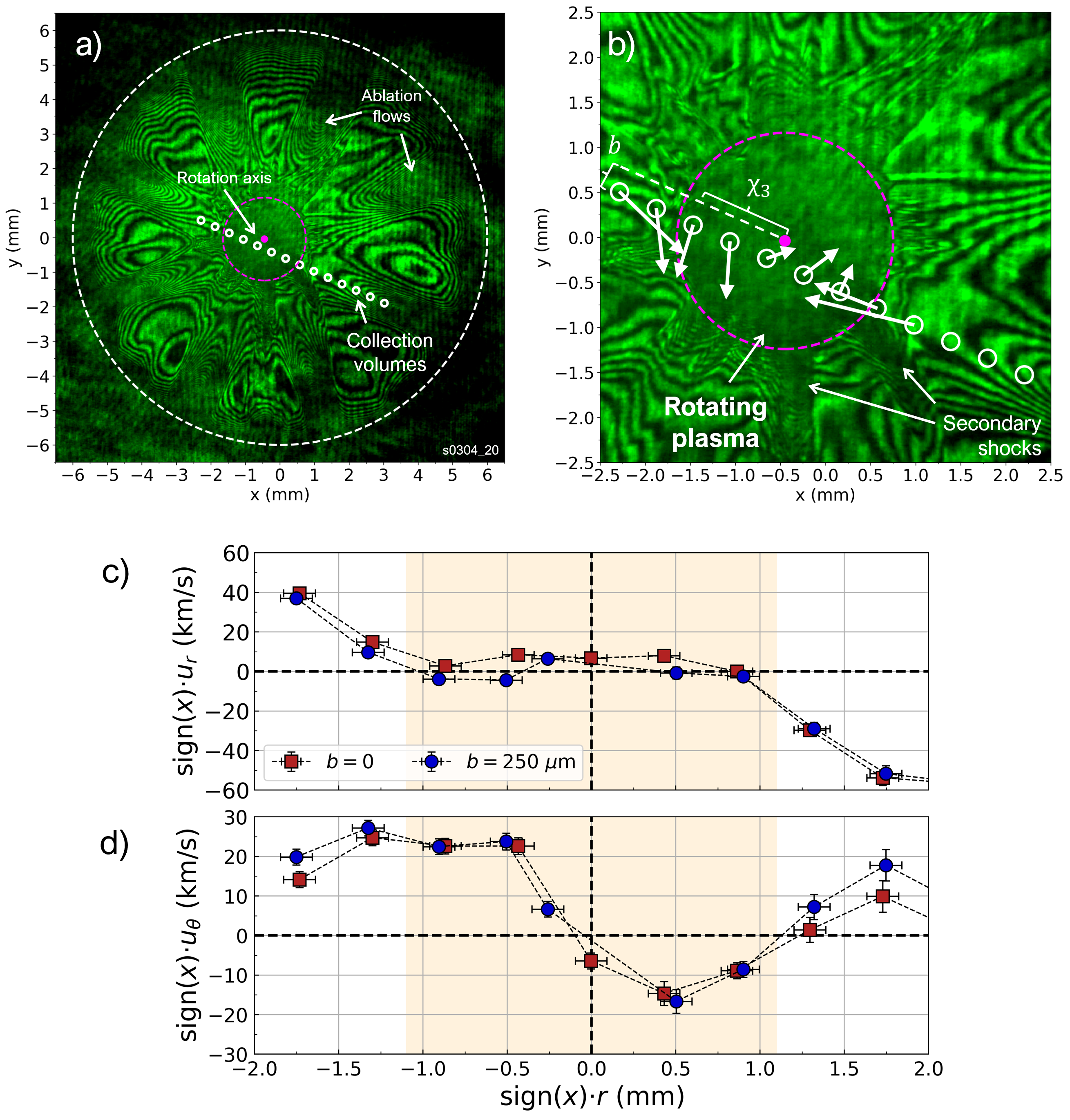}
    \caption{Horizontal flow velocity field on end-on interferometry imaging. a) Full end-on laser interferometry (532 nm wavelength). Inner cathode diameter (dashed circular line), outer edge of rotating plasma (magenta dashed circle), plasma symmetry axis (solid magenta circle), and TS collection volumes (small white circles) are presented. b) Close-up to the rotating plasma region. Flow velocity field obtained from TS Doppler are indicated using arrows to scale. c) Calculated radial velocity for the inferred impact parameter ($b= 250$ $\mu$m, blue circles) and zero impact parameter (same as Fig. \ref{fig:Thomson} for reference. d) Azimuthal velocity.}
    \label{fig:impact parameter}
\end{figure*}

\subsubsection{Effect of the probe impact parameter on the velocity measurements} \label{subsec:impact parameter}

In general, the mapping between the flow velocity components $(u_\parallel, u_\perp) \longrightarrow (u_r, u_\theta)$ depends on the orthogonal distance between the probe beam and the rotation axis, mathematically defined by the impact parameter $b$. The transformation between flow velocity components accounting for the impact parameter is given by
\begin{align}
    u_r(\chi_i) = \text{sign}(\chi_i)\frac{\chi_iu_\parallel + b u_\perp}{r_i},&\quad u_\theta(\chi_i) = \text{sign}(\chi_i)\frac{\chi_iu_\perp - b u_\parallel}{r_i}, \label{eq:corrected velocity components}
\end{align}
where $\chi_i$ is the location of the collection volume $i$ along the laser probe path, $(u_\parallel, u_\perp)$ can be inferred from the Doppler shift using equations (\ref{eq:velocity components}), and
\begin{equation}
    r_i(\chi_i) = \sqrt{\chi^2_i +b^2},
\end{equation}
is the distance from the collection volume to the rotation axis. It is evident that the particular case where $b = 0$ yields the result from the previous section, i.e. $u_r = \text{sign}(\chi_i)\cdot u_\parallel$ and $u_\theta = \text{sign}(\chi_i)\cdot u_\perp$. Moreover, the correction is important only in the region where $\chi_i \lesssim b$.

To experimentally correct the location of the Thomson Scattering probe beam relative to the rotation axis in the transverse direction, we map the measurements onto our end on interferometry data. Fig. \ref{fig:impact parameter}a shows a full shot interferogram with the TS collection volumes drawn. A close-up view to the region of interested in presented in Fig. \ref{fig:impact parameter}b. The region where interferometry fringes are lost define the shell-like density structure of the rotating plasma, and its outer boundary is indicated in a dashed magenta circle. Moreover, the axis of such circle is indicated with a solid circle (indicated in panel a). Assuming the axis of plasma symmetry (i.e. where the center of mass of the system should be) corresponds to the rotation axis, and noticing that the white circles indicate the path of the probe beam, a value of the impact parameter $b = 250 \pm 50$ $\mu$m can be estimated. In addition, the horizontal flow velocity field evaluated on the collection volumes can be obtained by combining $(u_A,u_B)$, which is shown as white arrows in Fig. \ref{fig:impact parameter}b.

The result of the correction on the radial and azimuthal flow velocity components is shown in  Fig. \ref{fig:impact parameter}c and d, respectively. The result was discussed only schematically by \cite{Valenzuela-Villaseca2023} to show that the identification of a quasi-Keplerian rotation curve is not dependent on the impact parameter. However, here we present a direct measurement using a diagnostic with a line of sight orthogonal to the TS geometry. We find that if this correction is ignored, then the result indicates that there is a constant side-ways velocity, however after accounting for the impact parameter allows correctly interpreting that near the axis $u_\parallel$ corresponds to $u_\theta$ rather than $u_r$. Although the magnitude of the rotation velocity changes, this correction may be important to correctly determine rotation curves close to the axis.

\section{Discussion: comparison to the self-similar evolution of a thin rotating shell} \label{sec:discussion}

\begin{figure*}
    \centering
    \includegraphics[width=13cm]{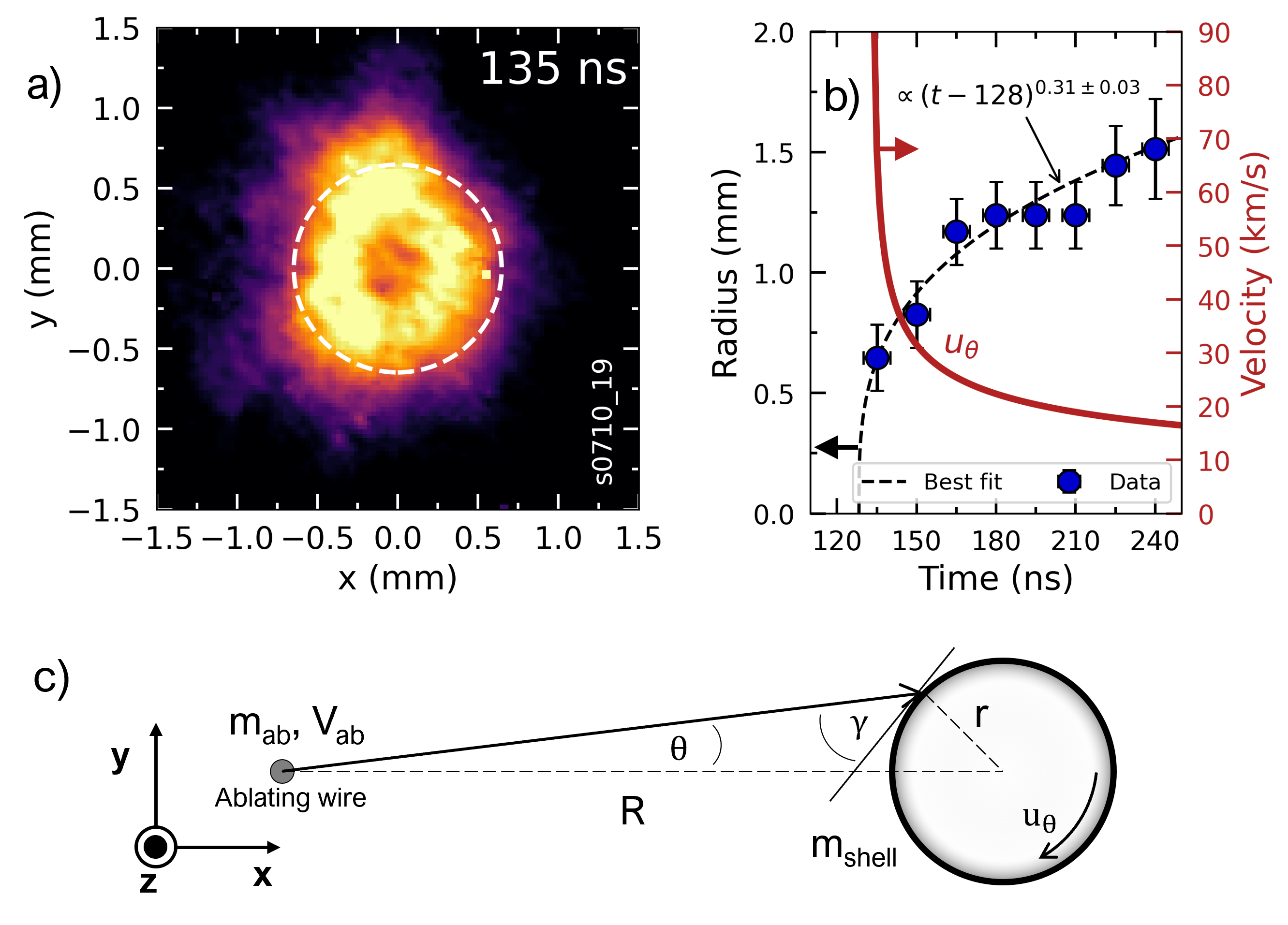}
    \caption{Analysis of rotating plasma evolution from optical self-emission imaging. a) Close up to rotating plasma region presented in Fig. 3. Dashed white circle indicates approximate full width at half maximum intensity. b) Evolution of outer and inner characteristic radii obtained for each image in the sequence. c) Schematic of ablation model. For simplicity, only one wire is presented rather than the $N$ wires considered in the full self-similar model (c.f. Fig. 1).}
    \label{fig:self similar}
\end{figure*}

The sequence of optical self-emission images allow tracking the radial evolution of the rotating plasma through time. As presented above, the plasma has a broadband self-emission distribution corresponding to a bright shell with reduced intensity on axis. Our goal is to use the information encoded in the radial dynamics to investigate the role of the ablation flows in the confinement and angular momentum injection in the experiment. In this section we derive a self-similar solution (meaning that the radius and rotation velocity at a given time $t_1$ can be scaled to a different time $t_2$) for a thin plasma shell driven and confined by the inflowing ablation flows. It is assumed that the shell-width $\delta$ and mean-radius $r$ of the plasma are such that $\delta /r \ll 1$ over their evolution. We compare the data with the model and find that the edge of the plasma follows the self-similar trajectory of the thin shell, and additionally, the measured rotation velocity is consistent with this simple mathematical description. To quantify the plasma radial dynamics, we have defined an inner and an outer characteristic plasma radii at half the height  of the intensity peak of the shell, as indicated in Fig. \ref{fig:self similar}a.

Fig. \ref{fig:self similar}b shows the evolution of the charcteristic plasma radius throughout the experiment. We observe that the plasma outer radius steadily grows in time. To calculate the radial growth of the overall rotating plasma, we use an approach similar to \cite{Ampleford2008}, shown in Fig. \ref{fig:self similar}c. We consider the ablation of N wires described by the rocket model \cite{Lebedev2001a} which continuously inject mass on a thin shell which retains all the mass accumulated during the ablation. We assume that the plasma stagnated in the shell remains confined by the counter-balance of the centrifugal force and momentum flux, i.e.
\begin{equation}
    NV_{ab}\sin{\gamma}\frac{d}{dt}m_{ab} = \frac{u_\theta^2}{r}m_{\text{shell}}, 
\end{equation}
where $V_{ab}$ is the rocket model ablation velocity, $m_{shell}$ is the shell’s total mass, $\gamma$ is the contact angle between the ablation flow and the shell, $m_{ab}$ is the ablated mass, $u_\theta$ is the rotation velocity of the shell and $r$ is the shell radius. In addition, we also assume that the angular momentum of each ablation flow is conserved during their propagation, therefore the velocity at which it reaches the rotating shell is given by
\begin{equation}
    u_\theta = \frac{R}{r}V_{ab}\sin{\theta}, \label{eq:ang momentum}
\end{equation}
where $R$ is the wire array radius. Finally, the mass of the shell is the accumulation of mass ablated from the wires, and therefore
\begin{equation}
    m_{\text{shell}} \sim N\cdot\frac{d}{dt}m_{ab}\cdot\left(t - t_{\text{ret}}\right),
\end{equation}
where $t_{\text{ret}} \sim R/V_{ab}$ is the retarded time that takes a plasma parcel travelling on an ablation flow from the wire to reach the shell (provided $R \gg r$). These three equations correspond to the conservation of linear momentum, angular momentum, and mass. Substituting them yields the self-similar evolution of the shell’s radius

\begin{equation}
    r(t) = \left[ \frac{\sin^2{\theta}}{\sin{\gamma}}R^2 V_{ab} \left( t - \frac{R}{V_{ab}} \right)  \right]^{1/3}. \label{eq: self similar radius}
\end{equation}

The data obtained from the optical self-emission images were fitted with a least-square method using a function of the form $A(t+t_0)^n$. The results show that the exponent in the experiment is $n = 0.30 \pm 0.03$, and therefore is consistent with the self-similar model. The fit predicts a formation time $t_0 = 128 \pm 10$ ns, consistent with the formation time $\sim 130$ ns observed from optical self-emission images. 

Further evidence supporting this result can be found by calculating the evolution of the shell’s characteristic rotation velocity. The self-similar expression for this velocity can be computed by replacing equation \ref{eq: self similar radius} into \ref{eq:ang momentum}, yielding

\begin{equation}\label{eq:rotation velocity evolution}
    u_\theta(t) = \left[\sin{\theta} \sin{\gamma} \frac{R V_{ab}^2}{t - R/V_{ab}} \right]^{1/3}.
\end{equation}

Fig. \ref{fig:self similar}b presents the calculated rotation velocity evolution on the right-hand-side vertical axis. The self-similar solution has a singularity at $t=t_{\text{ret}}$ which corresponds to the formation time which is not described by the model. However, as the ablation flows feed mass to the rotating plasma, and its radius increases, its rotation velocity rapidly decreases due to the increasing moment of inertia. Although the scaling $u_\theta \sim t^{1/3}$ implies a monotonically decreasing rotation velocity with time, in practice the curve settles at $u_\theta \lesssim 20$ km/s, which is approximately what was measured experimentally.

The self-similar solution has good agreement with the radial evolution and the characteristic rotation velocity observed in the experiments. We notice that it does not assume any form of magnetic confinement, but rather it is all provided by the ram pressure of the ablation flows. A designer can use these equations to estimate the plasma size and rotation velocity given an experimental configuration by using the magneto-static equations (\ref{eq:radial}) and (\ref{eq:azimuthal}) to calculate $\theta$ without the need of computationally expensive simulations.

An interesting prediction of equations (\ref{eq: self similar radius}) and (\ref{eq:rotation velocity evolution}) is that they have reciprocal time scaling. Hence, the magnitude of inertial forces $\sim ru_\theta = \ell $ is a conserved quantity. This implies that the fluid and magnetic Reynolds numbers, both $\propto r u_\theta$, only depend on material properties such as density and temperature. A nice property of the Plasma Couette Experiment \citep{Collins2012} is that they can pump down the chamber during the discharge to vary the magnetic Prandtl number Pm $=\nu/\eta$ (the ratio of viscosity to magnetic diffusivity). This parameter is crucial in setting the hierarchy of scales in MRI turbulence, dynamo, and angular momentum transport \citep{Schekochihin2004,Lesur2021}. Current pulsed-power experiments can drive low magnetic Prandtl plasmas, with Pm $\sim 10^{-2}$. Speculatively, in a sufficiently long duration RPX experiment with the right plasma composition, radiative cooling could be used to slowly change the value of Pm relative to unity and study its effect in turbulence in a differentially rotating, quasi-Keplerian background.

\section{Conclusions}\label{sec:conclusion}

We have presented a detailed characterization of the structure and dynamics of rotating plasmas driven on the RPX pulsed-power platform. This report complements a previous publication by \cite{Valenzuela-Villaseca2023}, as follows:\\

\begin{itemize}
    \item[1.] We have used laser interferometry to investigate the ablation dynamics of the wire array Z-pinch. The net effect of the radial magnetic field is to re-direct the ablation flows slightly off-radially by an angle $\theta$. Using a magneto-static model, we estimated $\theta=4^{\circ}$, which is in good agreement with experimental results.
    \item[2.] Side-on measurements of the axial jet show that the hollow density profile is sustained as it propagates vertically upwards. This shows that rotation is dynamically significant and provides a centrifugal barrier. XUV images have shown that the plasma does not change significantly its morphology up until the shell-like implosion of the Z-pinch reaches the axis at $t=270$ ns. At that time we observe the axial jet to kink.
    \item[3.] We have measured directly the off-set distance of the TS probe with respect to the plasma symmetry axis. We have corrected the radial and azimuthal velocity measurements and shown that the radial velocity is zero within uncertainty, and that the inferred rotation velocity does not change significantly compared to our previous report.
    \item[4.] We have derived a self-similar solution describing the radial growth and velocity evolution of a thin, rotating plasma shell. The rotating plasma grows and slows down as it accretes mass and angular momentum from the ablation flows, which also provide radial confinement through ram pressure. The model shows that the rotation velocity gradually decreases as the plasma moment of inertia increases. However, this decay saturates at an estimated the velocity of $u_\theta \sim 20$ km/s, which is in agreement with direct TS measurements. We find that both the outer boundary of the plasma in the experiment and measured rotation velocity are consistent the self-similar trajectory. A straight forward application of this result is to couple this self-similar solution with the calculated magnetic field components presented earlier in the paper, to easily design alternative RPX configurations and generate plasmas with a given radius and rotation velocity, without the need for computationally expensive simulations.
\end{itemize}

Recently, \cite{Beresnyak2023} reported the observation of MRI in a Z pinch-like imploding geometry in the self-similar Mag Noh problem. Their more sophisticated solution allows to calculate the interior of the rotating plasma, which is beyond what our solution can provide. Future reports will investigate the dynamics of plasma under faster rotation, which we expect to exhibit larger diameter through which the plasma rotation curve can be more accurately probed due to better relative spatial resolution.

\section*{Acknowledgements}
 We are grateful to the Mechanical Workshop at the Imperial College Department of Physics, and in particular to David Williams, for hardware fabrication. We thank Philipp K.-S. Kempski for fruitful conversations about the equations presented in the discussion. We thank Thomas Varnish for writing some of the codes used in the analysis. Vicente Valenzuela-Villaseca is currently at the Department of Astrophysical Sciences, Princeton University, Princeton, New Jersey, USA. Francisco Suzuki-Vidal is currently at First Light Fusion Ltd., Yarnton, UK. Jack W. D. Halliday is currently with at the Department of Physics, Clarendon Laboratory, University of Oxford, Oxford, UK. Danny R. Russell is currently at the Technische Universitaet Muenchen, Forschungs Neutronenquelle Heinz Maier-Leibnitz,  Munich, Germany.

\section*{Funding}
This work was supported in part by NNSA under DOE Cooperative Agreement No DE-SC0020434 and DE-NA0003764. The work of Vicente Valenzuela-Villaseca was supported by the Imperial College President's PhD Scholarships and the Royal Astronomical Society.

\section*{Declaration of Interests}
The authors report no conflict of interest.

\bibliographystyle{jpp}

\bibliography{jpp-bib.bib}

\end{document}